# Moving Target Detection Method Based on Range-Doppler Domain Compensation and Cancellation for UAV Mounted Radar

Xiaodong Qu, Member, IEEE

Xiaolong Sun

Feiyang Liu

Hao Zhang

Shichao Zhong, Member, IEEE

Xiaopeng Yang, Senior Member, IEEE

*Abstract*—**Combining unmanned aerial vehicle (UAV) with through-the-wall radar can realize moving targets detection in complex building scenes. However, clutters generated by obstacles and static objects are always stronger and non-stationary, which results in heavy impacts on moving targets detection. To address this issue, this paper proposes a moving target detection method based on Range-Doppler domain compensation and cancellation for UAV mounted dual channel radar. In the proposed method, phase compensation is performed on the dual channel in range-Doppler domain and then cancellation is utilized to achieve roughly clutters suppression. Next, a filter is constructed based on the cancellation result and the raw echoes, which is used to suppress stationary clutter furthermore. Finally, mismatch imaging is used to focus moving target for detection. Both simulation and UAV-based experiment results are analyzed to verify the efficacy and practicability of the proposed method.**

## I. INTRODUCTION

THE UAV mounted through-the-wall radar is capable of detecting hidden moving targets within buildings, which is employed extensively in urban combat, anti-terrorism, et al.[1][2]. Due to the complexity of the building environment, moving target detection has great technical challenges, such as the clutter caused by obstacles is always much stronger than signals from the moving targets. In addition, the cancellation performance degrades heavily due to unstable UAV moving trajectory[3].

To suppress the clutters, multi-channel radar was always adopted. Classical methods mainly included Along-track Interferometry (ATI)[4], Displaced Phase Center Antenna (DPCA)[5], Space Time Adaptive Processing (STAP)[6], and so on. The core idea of these classical methods was that the phases of moving targets were different across multi-channels while those of stationary targets kept unchanged. For example, the joint-pixel DPCA proposed by Yang [7] and the E-DPCA method proposed by Cerutti [8] could improve the detection performance of DPCA algorithm in specific scenes. Later, Wang [9] proposed a novel detection method, which could improve moving targets detection ability under the conditions of low channel correlation and low degree of alignment. In addition, a two-step clutter suppression method based on GO-DPCA and Local STAP was proposed [10], in which GO-DPCA was utilized to detect moving target and Local STAP to suppress stationary clutter. The method applied GO-DPCA and Local STAP together, which saved a lot of computation and improved the output SNR of DPCA. In [11], a generalized adaptive matched filter with multidimensional search was proposed to detect maneuvering targets using airborne radar systems, in which the computational complexity was reduced to great extent. In [12], a fast and robust STAP method based on sparse recovery technology was proposed, which reduced the excessive calculation burden of STAP algorithm and improved its suppression performance for non-uniform clutter. However, these methods mentioned above were influenced by the channel spacing, sampling interval, and platform speed, resulted in performance deterioration during practical applications.

Then, robust principal component analysis (RPCA) technique based on low-rank and sparse decomposition was proposed[13]. In [14], RPCA-Net was proposed to iteratively generate hyperparameters, which were applied into RPCA to realize moving targets detection. The NSS-RPCA was proposed in [15] to improve the suppression performance for non-heterogeneous environmental clutter, which realized moving targets detection effectively in non-uniform clutter. In additional, the NRPCA [16] was proposed to deal with the

Manuscript received XXXXXXXX XXth, 2024; revised XXXXXXXX XXth, 2024; accepted XXXXXXXX XXth, 2024.

Date of publication XXXXXXXX XXth, 2024; date of current version XXXXXXXX XXth, 2024. This work was supported in part by the National Natural Science Foundation of China under Grant 62101042. This work was also supported in part by Beijing Institute of Technology Research Fund Program for Young Scholars under Grant XSQD-202205005. (Corresponding author: Xiaopeng Yang.)

Xiaodong Qu, Xiaolong Sun, and Feiyang Liu are with the School of Information and Electronics, Beijing Institute of Technology, Beijing 100081, China, and with the Key Laboratory of Electronic and Information Technology in Satellite Navigation, Beijing Institute of Technology, Beijing 100081, China (e-mail: xdqu@bit.edu.cn; 3120220731@bit.edu.cn; feiyangliu0618@163.com).

Hao Zhang, Shichao Zhong, and Xiaopeng Yang, are with the School of Information and Electronics, Beijing Institute of Technology, Beijing 100081, China, and also with the Jiaxing Research Center of Beijing Institute of Technology, Jiaxing 314000, China (email: 3220220662@bit.edu.cn; zhongshichao16@163.com; xiaopengyang@bit.edu.cn;).

problem caused by non-sparse moving target echoes. In [17], a low-rank and sparse decomposition algorithm for 3D radar data (frequency, space, and time) was proposed to model wall clutter suppression and target detection as an optimization problem, which was able to separate walls, stationary targets, and moving targets simultaneously. However, the methods based on low-rank and sparse decomposition had higher requirements on the number of channels.

In addition, there were also some useful methods used for moving target detection. For example, UFJSR-Net[18] was proposed to improve the detection performance of moving targets, which mainly addressed the lack of adaptivity and difficulty in selecting appropriate hyperparameters in traditional methods. In [19], the authors proposed ground moving target detection and refocusing techniques based on dual-channel transceivers and up-down linear frequency modulated signals. To deal with indoor human target detection problems based on through-the-wall radar, a postprocessing method is proposed in [20] to detect multiple adjacent human targets. In [21], for MIMO radar system, a neural network was introduced to reduce multi-channel mutual interference, increasing significantly target detection accuracy. In [22], a multi-frame detection method based on image entropy was utilized to localize regions of interest.

When the moving target is behind an obstacle and very close to the UAV-mounted radar, the approximation used by ordinary moving target detection methods is no longer applicable. In additional, the sparsity of moving targets usually can't meet the requirement of RPCA method. Therefore, an indoor moving target detection method based on Range-Doppler domain phase compensation, cancellation, filtering, and mismatch imaging, is proposed using UAV-mounted dual channel radar.

The structure of this article is as follows. Echo model for moving target by multichannel radar is described in Section II. In Section III, clutter suppression methods are presented in detail. Section IV analyzes the results and verifies the effectiveness. Finally, Section V provides a concise conclusion.

## II. MOVING TARGET ECHO MODELING FOR UAV-MOUNTED MULTI-CHANNEL RADAR

Fig. 1 shows the scene of the UAV-mounted radar and the moving target inside the building. In the Cartesian coordinate system, $X$ axis is the azimuthal direction along which the UAV moves, the $Y$ axis corresponds to range direction, and the $Z$ axis is altitude. The origin is the location of the first receiving antenna projection to the ground, and the location of the target is $(x_0, y_0)$.

Assuming that the radar contains one transmitter and $N$ receivers, and linear frequency modulated (LFM) wave is used. The radar echo of moving target is the delayed edition of the transmitting signal. After demodulation, the received echo of the $n$ th channel $s_n(\tau, \eta), n = 1, 2, \cdots, N$ is as follows:

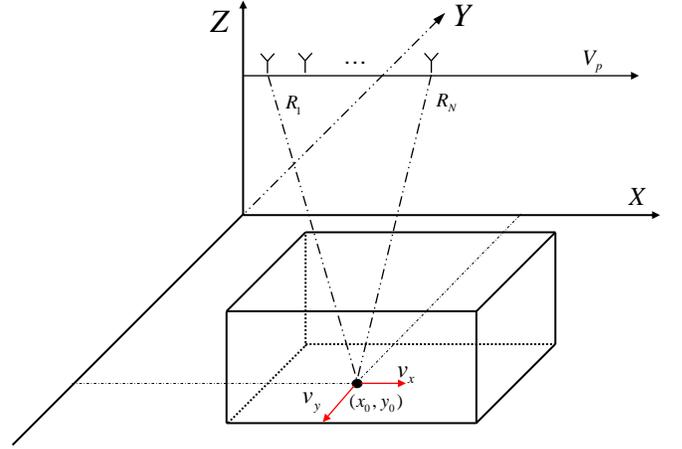

Fig. 1. Schematic diagram for UAV-mounted through-the-wall radar detecting moving target inside building.

$$
\begin{aligned}
s_n(\tau, \eta) = {} & A_0 w_r(\tau - R_n(\eta)/c) w_a(\eta - \eta_c) \\
& \times \exp\{-j 2\pi f_0 R_n(\eta)/c\} \\
& \times \exp\{j\pi K_r(\tau - R_n(\eta)/c)^2\}
\end{aligned} \tag{1}
$$

where $A_0$ is a complex constant, $\tau$ is the fast time, $\eta$ is the slow time, $\eta_c$ is the beam center azimuth deviation time. $w_r$ and $w_a$ are the envelopes of range and azimuth, respectively. $R_n(\eta)$ is the two-way length from transmitter to the $n$ th receiver.

The phase of the moving target is directly determined by $R_n(\eta)$, which can be expanded into the following expression:

$$
\begin{aligned}
R_n(\eta) = {} & \sqrt{(x_0 + v_x\eta - v_p\eta - d_{tr})^2 + (y_0 + v_y\eta)^2 + H^2} \\
& + \sqrt{(x_0 + v_x\eta - v_p\eta + d_n)^2 + (y_0 + v_y\eta)^2 + H^2} \\
& + R_{wall}
\end{aligned} \tag{2}
$$

where $v_p$ is the velocity of platform and $v_x, v_y$ are the velocity of moving target. $H$ is the altitude at which the UAV flies. $d_n = (n-1)d$ is the distance from the $n$ th receiver to the first receiver. $d_{tr}$ is the distance between the transmitter antenna and the first receiver. $d$ is the distance between the adjacent receivers. $R_{wall}$ is the extra length caused by wall and $R_{wall} = 2d_w\left(\sqrt{\varepsilon_r} - 1\right)$, in which $d_w, \varepsilon_r$ are the thickness and relative dielectric constant of the wall, respectively. Notably, in free space, the term $R_{wall} = 0$.

Then, the term $R_n(\eta)$ is expanded:

$$
R_n(\eta) \approx a_{0\_n} + a_{1\_n}\eta + a_{2\_n}\eta^2 \tag{3}
$$

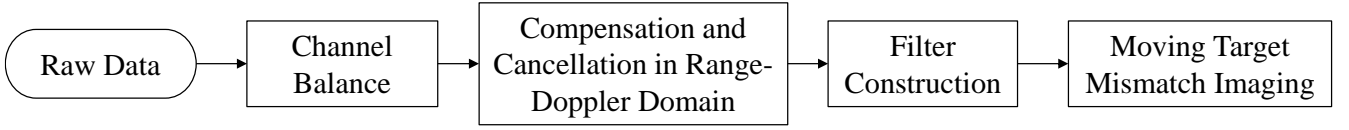

Fig. 2. Flow chart for moving target detection.

Assume that $R_0 = \sqrt{x_0^2 + y_0^2 + H^2}$, $v_r = x_0 v_x + y_0 v_y$, and the coefficients $a_0, a_1, a_2$ are:

$$a_{0\_n} = 2R_0 + \frac{d_{tr}^2}{2R_0} + \frac{x_0}{R_0}(d_n - d_{tr}) + \frac{d_n^2}{2R_0} + R_{wall}$$

$$a_{1\_n} = \frac{2(v_r - x_0 v_p) + (v_x - v_p)(d_n - d_{tr})}{R_0} \quad (4)$$

$$a_{2\_n} = \frac{(v_x - v_p)^2 + v_y^2}{R_0}$$

Combining eq.(3) and eq.(4), the echo signal $s_n(\tau, \eta)$ can be rewritten as:

$$\begin{aligned} s_n(\tau, \eta) = &A_0 w_r(\tau - R_n(\eta)/c) w_a(\eta - \eta_c) \\ &\times \exp\{j\pi K_r(\tau - R_n(\eta)/c)^2\} \\ &\times \exp\{-j\frac{2\pi}{\lambda}(a_{0\_n} + a_{1\_n}\eta + a_{2\_n}\eta^2)\} \end{aligned} \quad (5)$$

The reference signal used for pulse compression along fast time axis is:

$$s_{r\_ref}(\tau) = w_r(\tau/T_p)\exp(-j\pi K_r \tau^2) \quad (6)$$

And then, the signal after compression becomes:

$$\begin{aligned} s_{n\_rc}(\tau, \eta) &= IFFT_\tau\left\{FFT_\tau\left[s_n(\tau, \eta)\right] FFT\left[s_{r\_ref}(\tau)\right]\right\} \\ &= A_0 p_r(\tau - R_n(\eta)/c) w_a(\eta - \eta_c) \\ &\times \exp\{-j\frac{2\pi}{\lambda}(a_{0\_n} + a_{1\_n}\eta + a_{2\_n}\eta^2)\} \end{aligned} \quad (7)$$

where $p_r(\tau)$ is the *sinc* function.

## III. PROPOSED CLUTTER SUPPRESSION METHOD

The flow chart is shown in Fig.2. Channel balance is first applied on the raw data to obtain better consistency across different channels. Then, compensation and cancellation methods in range-Doppler domain are applied to the balanced data to remove stationary clutter. Next, a filter in time domain which is constructed based on cancellation results and raw echoes is used to suppress stationary clutters furthermore.

Finally, mismatch imaging is performed to obtain the images of moving targets for detection.

### A. Channel Balance

A multichannel array system is essentially a combination of several independent systems, in which each receiving channel can be regarded as a subsystem. Different subsystems differ from each other in terms of frequency response and amplitude response, which will impose heavy impacts on the final performance [23], [24]. Therefore, energy normalization is used to preprocess the received echo, which can be expressed as:

$$s_{n\_cb}(\tau, \eta_k) = \frac{s_{n\_rc}(\tau, \eta_k)}{\sum_\tau s_{n\_rc}(\tau, \eta_k)^2} \quad (8)$$

### B. Compensation and cancellation in Range-Doppler domain

Then, the balanced signal is transformed to the Range-Doppler domain using the principle of stationary phase:

$$\begin{aligned} s_{n\_cb}(\tau, f_\eta) = &A_1 p_r(\tau - R_n(f_\eta)/c) W_a(f_\eta - f_{\eta c}) \\ &\times \exp\{j\phi_n(f_\eta)\} \end{aligned} \quad (9)$$

where $W_a(f_\eta - f_{\eta c})$ is the frequency domain form of the azimuthal envelope $w_a(\eta - \eta_c)$.

The phase term $\phi_n(\eta)$ in time domain is:

$$\phi_n(\eta) = -\frac{2\pi}{\lambda}(a_{0\_n} + a_{1\_n}\eta + a_{2\_n}\eta^2) \quad (10)$$

and thus, the phase term $\phi_n(f_\eta)$ in frequency domain is derived as follows:

$$\begin{aligned} \frac{d\phi_n}{d\eta} &= -\frac{2\pi}{\lambda}(a_{1\_n} + 2a_{2\_n}\eta) - 2\pi f_\eta = 0 \\ f_\eta &= -\frac{a_{1\_n} + 2a_{2\_n}\eta}{\lambda} \\ \eta &= \frac{-\lambda f_\eta - a_{1\_n}}{2a_{2\_n}} \end{aligned} \quad (11)$$

Combining (10) and (11), the expression $\phi_n(f_\eta)$ is:

$$\phi_n(f_\eta) = \frac{2\pi}{\lambda}\left[\left(-a_{0\_n} + \frac{a_{1\_n}^2}{4a_{2\_n}}\right) + \frac{\lambda a_{1\_n}}{2a_{2\_n}}f_\eta + \frac{\lambda^2}{4a_{2\_n}}f_\eta^2\right] \quad (12)$$

In order to eliminate stationary clutter, the phase of stationary targets needs to be analyzed. From eq. (4), when the target is stationary, its velocity satisfies $v_x = 0, v_y = 0$, then the coefficients $a_{0\_n}, a_{1\_n}, a_{2\_n}$ can be simplified as follows:

$$a_{0\_n} = 2R_0 + \frac{d_{tr}^2}{2R_0} + \frac{x_0}{R_0}(d_n - d_{tr}) + \frac{d_n^2}{2R_0} + R_{wall}$$

$$a_{1\_n} = \frac{-2x_0 v_p - (d_n - d_{tr})v_p}{R_0} \qquad (13)$$

$$a_{2\_n} = \frac{v_p^2}{R_0}$$

In the Range-Doppler domain, there exists a certain phase difference across different channels for stationary point targets. The difference is denoted as $\Delta\phi_n(f_\eta) = \phi_n(f_\eta) - \phi_{n-1}(f_\eta)$, which is simplified to be:

$$\Delta\phi_n(f_\eta) = \frac{2\pi}{\lambda}\left(\frac{-(2n-1)d^2 - 2d_{tr}d}{4R_0}\right) + \pi f_\eta\left(-\frac{d}{v_p}\right) \qquad (14)$$

Thus, in the Range-Doppler domain, the following relationship exists between its adjacent channels:

$$s_{n\_cb}(\tau, f_\eta) = s_{n-1\_cb}(\tau, f_\eta) \times \exp\{j\Delta\phi_n(f_\eta)\} \qquad (15)$$

Since there is no slow time offset in the Range-Doppler domain, the pixel-by-pixel phase cancellation can be performed directly across the multi-channel data after phase compensation. The advantage is that there is no need to perform alignment operation in the time-domain which is limited by the distance between antennas and cannot be accurately aligned without sub-precision interpolation. Then, the echo after phase compensation and cancellation is expressed as follows:

$$s_{n-1\_cs}'(\tau, f_\eta) = s_{n-1\_cb}(\tau, f_\eta) \\ -s_{n\_cb}(\tau, f_\eta) \times \exp\{-j\Delta\phi_n(f_\eta)\} \qquad (16)$$

By using inverse Fast Fourier Transform, in the time domain, the expression is:

$$s_{n-1\_cs}(\tau, \eta) = ifft\left(s_{n-1\_cs}'(\tau, f_\eta)\right) \qquad (17)$$

The effect of phase compensation and cancellation is also analyzed for moving targets. From eq.(4), the phase difference $\Delta\phi_n'(f_\eta)$ between two adjacent channels is:

$$\Delta\phi_n'(f_\eta) = \frac{2\pi}{\lambda} \cdot \frac{(v_x - v_p)^2\left(-(2n-1)d^2 - 2d_{tr}d\right)}{4\left((v_x - v_p)^2 + v_y^2\right)R_0}$$

$$+ \frac{2\pi}{\lambda} \cdot \frac{dv_y\left(4(v_x - v_p)y_0 - (4n-2)dv_y - 4x_0 v_y\right)}{4\left((v_x - v_p)^2 + v_y^2\right)R_0} \qquad (18)$$

$$+ \pi f_\eta \frac{(v_x - v_p)d}{(v_x - v_p)^2 + v_y^2}$$

Specially, when the target moves along the range direction, i.e. $v_x = 0$, the phase difference $\Delta\phi_n'(f_\eta)$ is:

$$\Delta\phi'(f_\eta) = \frac{2\pi}{\lambda} \cdot \frac{v_p^2\left(-(2n-1)d^2 - 2d_{tr}d\right)}{4\left(v_p^2 + v_y^2\right)R_0}$$

$$- \frac{2\pi}{\lambda} \cdot \frac{v_y^2\left((4n-2)d^2 + 4x_0 d\right) + 4dy_0 v_y v_p}{4\left(v_p^2 + v_y^2\right)R_0} \qquad (19)$$

$$+ \pi f_\eta \frac{-v_p d}{v_p^2 + v_y^2}$$

Obviously, using eq.(14) and eq.(16), the magnitude of stationary clutter is zero after cancellation in theory. By contrast, the most energy of moving target is reserved because of the additional phase change caused by the target motion, indicated by eq.(18) and eq.(19).

## C. Filter Construction

To further suppress clutter, a filter is constructed based on the ratio of $s_{n-1\_cs}(\tau, \eta)$ and $s_{n-1\_cb}(\tau, \eta)$ is expressed as follows:

$$f_{ratio}(\tau, \eta) = \begin{cases} 0, & \dfrac{\left|s_{n-1\_cs}(\tau, \eta)\right|}{\left|s_{n-1\_cb}(\tau, \eta)\right|} < \kappa \\ 1, & \dfrac{\left|s_{n-1\_cs}(\tau, \eta)\right|}{\left|s_{n-1\_cb}(\tau, \eta)\right|} \geq \kappa \end{cases} \qquad (20)$$

where, $\kappa$ is preset hyper parameter, which is determined by radar parameters $d, \lambda, v_p$ and target parameters $y_0, R_0, v_y$.

Based on the discussion above, the final results after stationary clutter suppression are obtained using:

$$s_{n-1\_STS}(\tau, \eta) = s_{n-1\_cs}(\tau, \eta) \cdot f_{ratio}(\tau, \eta) \qquad (21)$$

## D. Moving Target Mismatch Imaging

In theory, eq.(21) only includes the echoes of moving targets. Thus, to achieve better imaging results, it is of great benefits to analyze the echoes from moving targets $s_{n\_cb}'(\tau, \eta)$ and seek proper match filter function in azimuth. The echo after demodulation and compression is expressed as:

$$s'_{n\_cb}(\tau,\eta) = A_1 p_r(\tau - R_n(\eta)/c) w_a(\eta - \eta_c)$$
$$\times \exp\{-j\frac{2\pi}{\lambda} R_n(\eta)\} \quad (22)$$

where $R_n(\eta) = a_{0\_n} + a_{1\_n}\eta + a_{2\_n}\eta^2$.

The signal in the Range-Doppler domain is:

$$s'_{n\_cb}(\tau,f_\eta) = A_1 p_r(\tau - R_n(f_\eta)/c) W_a(f_\eta - f_{\eta c})$$
$$\times \exp\{j\phi_n(f_\eta)\} \quad (23)$$

The range envelope $R_n(f_\eta)$ is denoted as:

$$R_n(f_\eta) = a_{0\_n} - \frac{a_{1\_n}^2}{4a_{2\_n}} + \frac{\lambda^2}{4a_{2\_n}} f_\eta^2 \quad (24)$$

At the zero Doppler moment, the relevant parameters are expressed as follows:

$$a_{0\_n} = 2R_0 + \frac{d_{tr}^2}{2R_0} + \frac{d_n^2}{2R_0} + R_{wall}$$
$$a_{1\_n} = \frac{2v_r + (v_x - v_p)(d_n - d_{tr})}{R_0} \quad (25)$$
$$a_{2\_n} = \frac{(v_x - v_p)^2 + v_y^2}{R_0}$$

The range migration needs to be corrected is given by the third term of (24):

$$\Delta R_n(f_\eta) = \frac{\lambda^2 R_0}{4\left((v_x - v_p)^2 + v_y^2\right)} f_\eta^2 \quad (26)$$

After range cell migration correction, the signal changes to the following form:

$$s'_{n\_cb}(\tau,f_\eta) = A_1 p_r(\tau - \left(a_{0\_n} - \frac{a_{1\_n}^2}{4a_{2\_n}}\right)/c) W_a(f_\eta - f_{\eta c})$$
$$\times \exp\{j\phi_n(f_\eta)\} \quad (27)$$

At this point, the range envelope is independent of the slow time and the signal energy is concentrated at the shortest distance.

To image the target in the azimuth direction, the phase term in the (27) needs to be further analyzed, which is:

$$\phi_n(f_\eta) = \frac{2\pi}{\lambda}\left[\left(-a_{0\_n} + \frac{a_{1\_n}^2}{4a_{2\_n}}\right) + \frac{\lambda a_{1\_n}}{2a_{2\_n}} f_\eta + \frac{\lambda^2}{4a_{2\_n}} f_\eta^2\right] \quad (28)$$

then, the following azimuthal matched filter can be constructed:

$$H_{az}(f_\eta) = \exp\left(-j\frac{2\pi}{\lambda}\left(\frac{\lambda a_{1\_n}}{2a_{2\_n}} f_\eta + \frac{\lambda^2}{4a_{2\_n}} f_\eta^2\right)\right)$$
$$= \exp\left(-j\frac{2v_r + (v_x - v_p)(d_n - d_{tr})}{(v_x - v_p)^2 + v_y^2}\pi f_\eta\right) \quad (29)$$
$$\times \exp\left(-j\frac{\pi\lambda R_0}{2\left((v_x - v_p)^2 + v_y^2\right)} f_\eta^2\right)$$

Since the velocity of the moving target is unknown, it is not possible to construct an accurate azimuthal filter. However, in the detection scenario, the target of interest is limited to human beings. Thus, the velocity ranges from $0.3m/s$ to $1m/s$, so a mismatch filter can be constructed. If the azimuthal filter is constructed by using platform velocity alone without considering the velocity of the moving target, the filter will be a matched filter for stationary clutter and a mismatch filter for moving targets, which will be unfavorable for the detection of moving targets. On the contrary, if the velocity of the moving target is taken into account, the stationary clutter will be defocused while the focusing effect of the moving target will be improved, thus facilitating the detection of the moving target.

The final imaging result is:

$$s_{rd}(\tau,\eta) = ifft\left(fft\left(s_{n-1\_STS}(\tau,\eta)\right)\cdot H_{az}(f_\eta)\right) \quad (30)$$

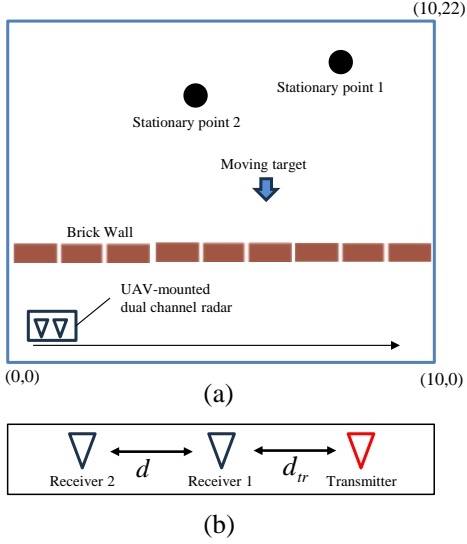

Fig. 3. (a) Simulation scene (b) Radar array

**TABLE I**
**SIMULATION PARAMETERS**

| Parameters | Value |
|---|---|
| Center frequency | 2.7 GHz |
| Bandwidth | 500 MHz |
| Pulse width | 440 us |
| PRF | 100 Hz |
| Distance between antennas | $0.06\,m$ , $0.4\,m$ |
| Position of static target | (6m,15m)/(4,14m) |
| Position of moving target | (5m,13m) |
| Velocity of moving target | $v_x = 0$ , $v_y = 1m/s$ |
| Velocity of platform | $v_p = 3m/s$ |
| Thickness of Wall | $0.24\,m$ |
| Relative permittivity of wall | 4 |

**TABLE II**
**IF RESULTS OF DIFFERENT METHODS**

| Scenes | RPCA | DPCA | Proposed Method | | | |
|---|---|---|---|---|---|---|
| | | | $v_y = 0m/s$ | $v_y = 0.3m/s$ | $v_y = 0.6m/s$ | $v_y = 0.9m/s$ |
| Simulation antennas space:0.06m | 5.17 dB | 37.67 dB | 45.12 dB | 48.68 dB | 50.36 dB | 53.71 dB |
| Simulation antennas space:0.40m | 5.76 dB | 10.09 dB | 41.16 dB | 43.13 dB | 45.11 dB | 46.32 dB |
| Experiment free space | 30.94 dB | 28.75 dB | 40.52 dB | 41.42 dB | 42.97 dB | 44.73 dB |
| Experiment through-the-wall | 25.12 dB | 21.77 dB | 37.70 dB | 41.50 dB | 43.62 dB | 45.33 dB |

## IV. SIMULATION AND EXPERIMENTAL RESULTS

Numerical simulations and field experiments in free space and through-the-wall scene are carried out in this section to verify the effectiveness. Moreover, a comparative analysis with RPCA and DPCA methods is presented. The improvement factor (IF) is used as the performance evaluation indicator. Its definition is:

$$IF = \frac{SCNR_{output}}{SCNR_{input}} = \frac{P_{out\_signal}/P_{out\_cluter}}{P_{in\_signal}/P_{in\_cluter}}$$

$$P_{clutter} = \frac{\sum_{M,L}\left|s_{rd}\left(\tau,\eta\right)\right|^2}{ML}$$

$$P_{signal} = \frac{\sum_{M',L'}\left|s_{rd}\left(\tau,\eta\right)\right|^2}{M'L'}$$

(31)

where $M$ , $L$ are the indices of clutters in imaging result and $M'$ , $L'$ are the indices of signal in imaging result.

### A. Numerical Simulation Results

Two simulation experiments are conducted, and the simulation scenario is shown in the Fig. **3**. The gprMax is used to generate the simulation echo data[25]. In the simulation, the UAV moves in the azimuth and the velocity of the UAV is 3 m/s. Two stationary points are used to simulate clutter and a point moving away from the radar along the range direction is used as motion target, whose velocity is 1m/s. The radar system has one transmitter and two receivers. The distances between antennas are set as 0.06m and 0.4m in two experiments, respectively. The center frequency is 2.7GHz with bandwidth of 500MHz. Other parameters are shown in the Table I.

The results are shown in Fig. 4 and the IFs are calculated in Table II. In the figure, (a)-(d) are the results when the distance between antennas is 0.06m and (e)-(h) are results when the distance between antennas is 0.4m. Fig. 4(a) and Fig. 4(e) are the directly imaging results using channel 2. Stationary point clutter is well focused and accurately positioned. However, the moving targets are poorly focused and the position of the moving target migrates a certain distance in the azimuthal direction compared with the initial position. The reason is that the human moved about 2m in range direction during the synthetic aperture time, resulting in that the imaging results are biased in range direction and azimuth direction. Fig. 4(b) and Fig. 4(f) are the results processed by RPCA using channel 2. RPCA is applied in the echo domain, and unable to eliminate the clutter due to that the low rank property of clutter does not hold, so that the stationary point clutter is still remained. Fig. 4(c) and Fig. 4(g) are the results processed by DPCA. When the distance between antennas is 0.06m, the clutter is suppressed well. However, when the distance between antennas is 0.4m, the clutter cannot be suppressed.

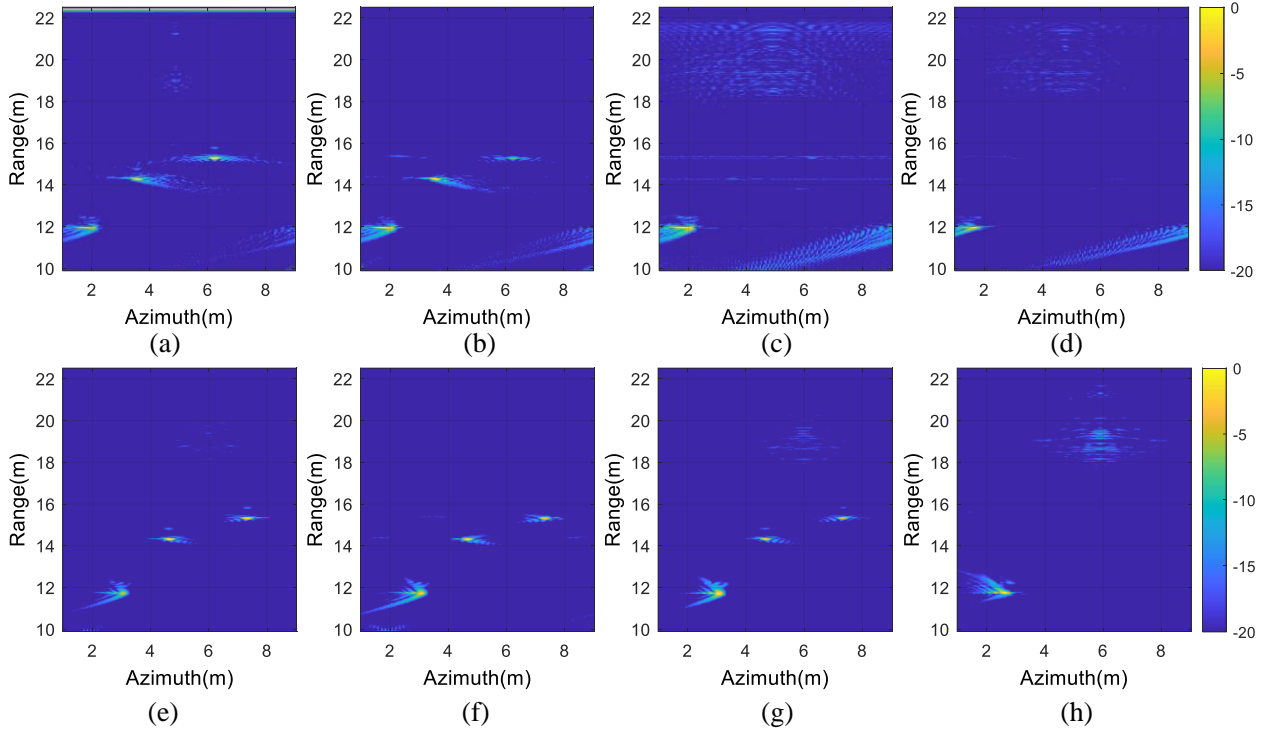

Fig. 4. Imaging results: (a)-(d) are the results when the distance between antennas is 0.06m and (e)-(h) are results when distance between antennas is 0.4m. (a) and (e) are the directly imaging results using channel 2. (b) and (f) are the results processed by RPCA using channel 2. (c) and (g) are the results processed by DPCA. (d) and (h) are the results processed by proposed method ($v_y = 1m/s$).

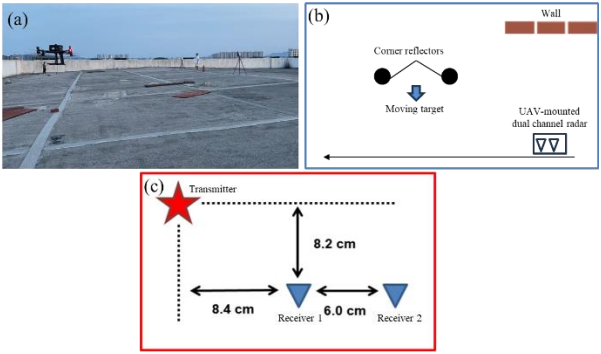

Fig. 5. Experiment in free space: (a) Experimental scenario (b) Schematic (c) Radar array

The reason is that when the distance between antennas is large, the phase of clutter across different channels cannot match. Fig. 4(d) and Fig. 4(h) are the results processed by the proposed method. It can be concluded that in both cases, no matter the distance between antennas is 0.06m or 0.4m, the clutter is suppressed successfully. As shown in TABLE II, the IFs of proposed methods are the highest. And when the velocity used to construct azimuth filter is closer to real velocity, the IF becomes larger.

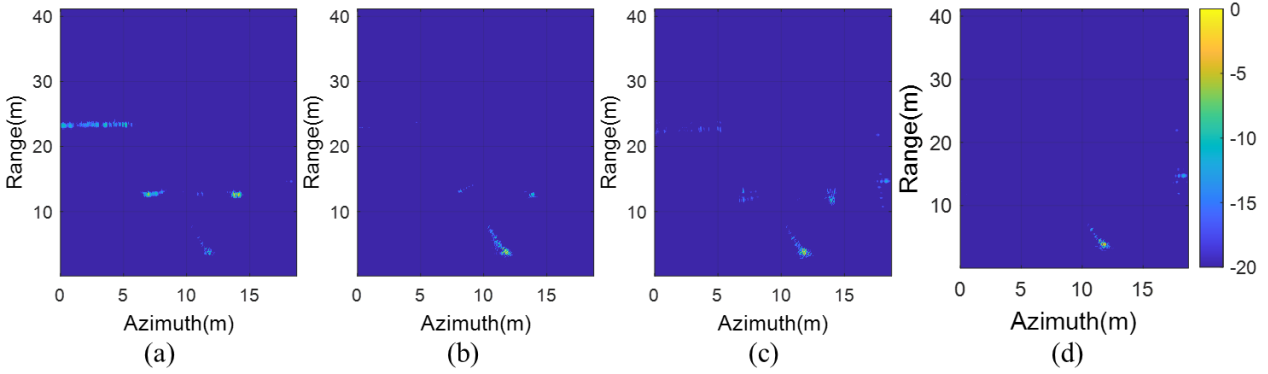

Fig. 6. Experimental results under free space: (a) is the directly imaging result using channel 2. (b) is the result processed by RPCA using channel 2. (c) is the result processed by DPCA. (d) is the result processed by proposed method ( $v_y = 1m/s$ ).

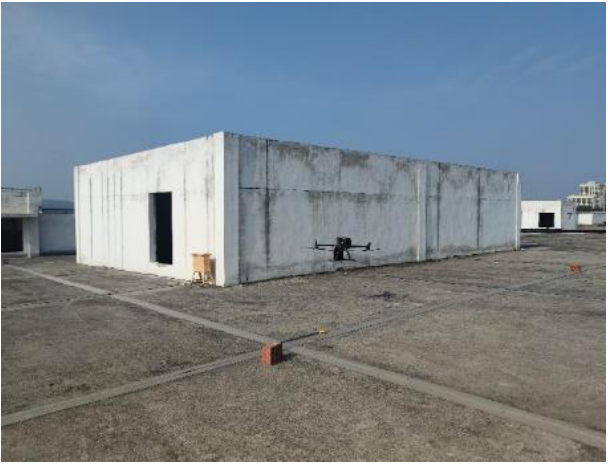

Fig. 7. Experimental scene for through-the-wall detection

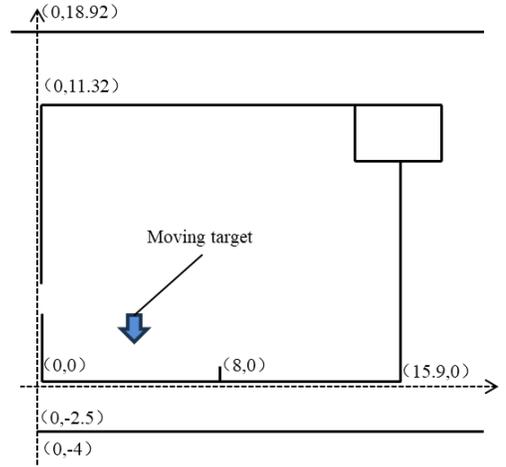

Fig. 8. Coordinate system for through-the-wall detection (origin is set at the corner of the wall)

### B. Experimental results under free space

The experimental scenario is shown in the Fig.5(a) and the schematic is shown in Fig.5(b). In this experiment, the human moves along the range direction. Two stationary corners reflectors are also used, and there is a section of building wall at the end. Thus, the clutters in the echoes mainly originate from corner reflectors and wall. The radar system has one transmitter and two receivers. The structure is shown in the Fig.5(c). The center frequency is 2.7GHz with bandwidth of 500MHz. Other parameters are shown in the Table III.

The results are presented in Fig. 6 and the IFs are also calculated in TABLE II. Fig. 6 (a) is the directly imaging result using channel 2. In the range direction, from near to far, the highlighted areas are moving target, two stationary corner reflectors and wall, respectively. Still, the stationary corner reflectors are well focused and accurately positioned, while the moving targets are poorly focused. Fig. 6 (b) is the result processed by RPCA using channel 2. Due to the strong low rank property of wall clutters, it can be eliminated by RPCA methods. However, the corner reflectors are still remained. Fig. 6 (c) is the results processed by DPCA, which shows that the clutters caused by corner reflectors and wall cannot be completely eliminated. Fig. 6 (d) is the result processed by the proposed method, showing that the clutters are suppressed successfully, and the moving target is well focused for detection. As shown in TABLE II, the proposed methods have the highest IFs. When the velocity used to construct azimuth filter becomes lager, which means that the velocity is closer to real velocity, the IF also increases.

### C. Experimental results for through-the-wall detection

We have conducted a through-the-wall detection experiment furthermore. The experiment scenario is shown in the Fig. 7. In the experiment, the human moves behind the wall and the UAV flies in the direction parallel to the wall. Fig. 8 is the top-view coordinate map corresponding to the real scene. The human moves within the scanning range, from (5,6) to (5,2). The UAV moves synchronously from (0, -2.5) to (8, -2.5). The radar system and parameters are the same as that in section B.

The experimental results are presented in Fig.9 and the IFs are also calculated in TABLE II. Fig. 9 (a) is the directly imaging result using channel 2, which only shows the image of wall. Due to the attenuation when penetrating the wall, the moving target is almost invisible. Fig. 9 (b) is the result processed by RPCA using channel 2. Due to the unstable motion trajectory of the UAV platform, the suppression performance of RPCA is affected, resulting in some residual wall clutter. Fig. 9(c) is the result processed by DPCA. As with the RPCA method, the DPCA method cannot completely eliminate wall clutter. Fig. 9 (d) is the result processed by the proposed method. The wall clutters are almost completely suppressed, while the moving target is well focused. The comparative analysis shows that the proposed method can notably improve the signal-to-clutter ratio, which is conducive for the detection task. As shown in TABLE II, using mismatch imaging methods improves the IFs and the performance of moving target imaging.

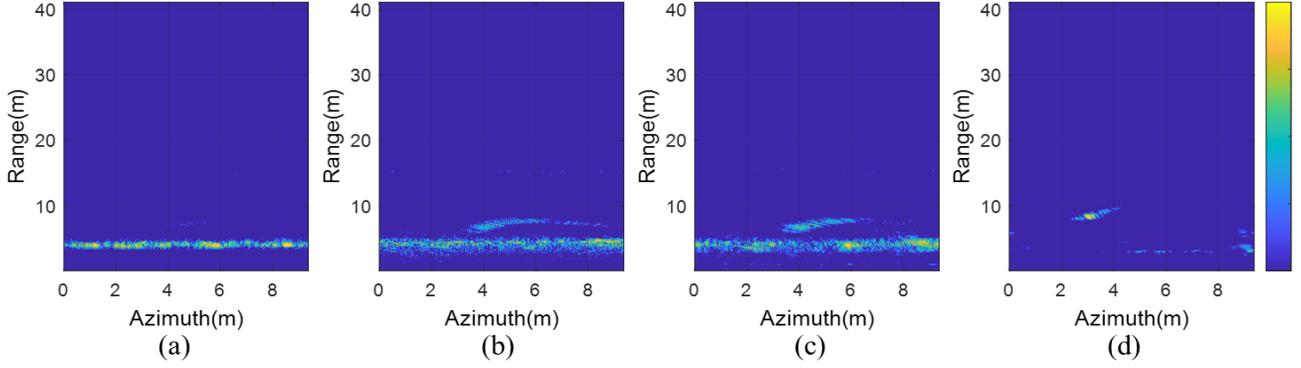

Fig. 9. Experimental results for through-the-wall detection: (a) is the directly imaging result using channel 2. (b) is the result processed by RPCA using channel 2. (c) is the result processed by DPCA. (d) is the result processed by proposed method ( $v_y = 0.9 m / s$ ).

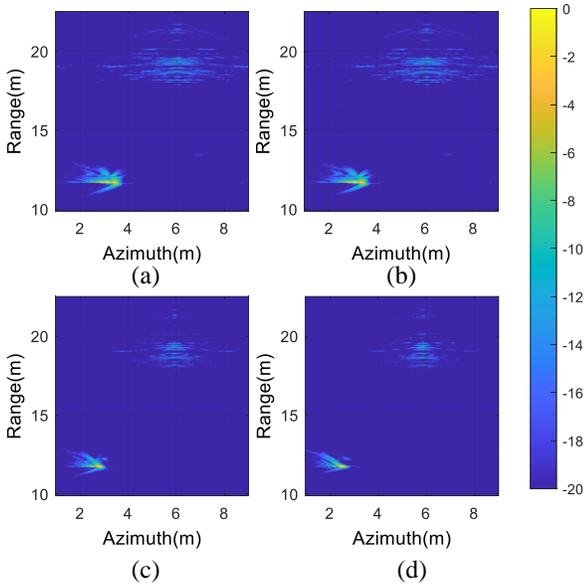

Fig. 10. Simulation results for different velocity of moving target used in constructing azimuth filter: (a) $v_y = 0$ m/s. (b) $v_y = 0.3$ m/s. (c) $v_y = 0.6$ m/s. (d) $v_y = 0.9$ m/s.

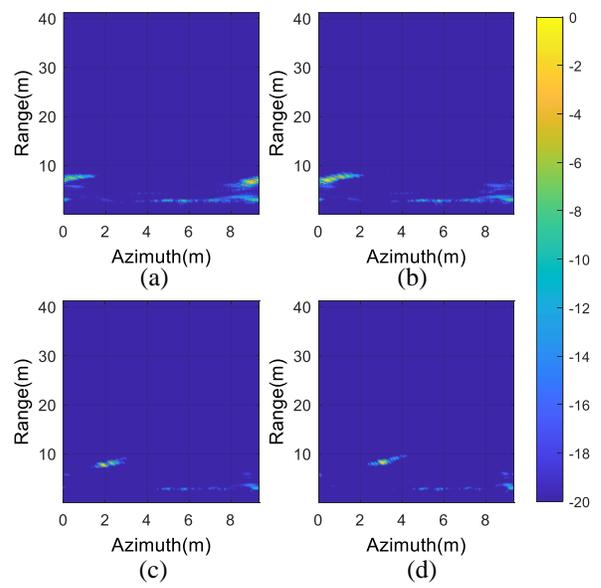

Fig. 11. Experimental results for different velocity of moving target used in constructing azimuth filter: (a) $v_y = 0$ m/s. (b) $v_y = 0.3$ m/s. (c) $v_y = 0.6$ m/s. (d) $v_y = 0.9$ m/s.

**D. The effect of velocity on constructing azimuthal filter**

Under the simulation conditions in Section A, the velocity value for constructing filter is changed to verify its influence on the imaging results. The results are shown in Fig.10 and the calculated IFs are shown in Table II. It can be seen that using mismatch filter is beneficial for moving target to be focused. And the closer the estimated speed is to the true value, the better the focusing effect will be.

Under the experiment conditions in Section C, the velocity value for constructing filter is changed. The results are shown in Fig.11 and the calculated IFs are also shown in Table II. When the velocity is 0, IF of the imaging result is 37.70 dB, and as the velocity value increases and approaches the true speed, the IF value then becomes larger. So, when constructing azimuth filter, taking velocity of moving target into consideration is beneficial to focus moving target and suppress clutter. In addition, the selection of the speed value is not troublesome, for the reason that any value chosen within the range of the target speed will achieve better results compared with not taking target's speed into consideration.

## V. CONCLUSION

This paper has proposed a method for detecting moving targets based on compensation and cancellation in range-Doppler domain. First, energy normalization was performed to balance the echoes from different channels. Then, the phase characteristics of moving targets and stationary clutters were analyzed in range-Doppler domain. Next, cancellation and filtering were performed in sequential to suppress clutters. Finally, azimuthal filter was constructed and used for mismatch imaging, which showed benefits for moving target detection. By processing and analyzing the measurements acquired by a UAV-mounted through-the-wall radar, the effectiveness and practicality of the method in detecting moving human inside buildings were verified. The results showed that the proposed method could effectively suppress clutters caused by static objects and walls, and significantly improve the detectability of moving targets.

## REFERENCES

[1] K. Wang, Z. Zeng and J. Sun, "Through-Wall Detection of the Moving Paths and Vital Signs of Human Beings," in *IEEE Geoscience and Remote Sensing Letters*, vol. 16, no. 5, pp. 717-721, May 2019.

[2] H. Li, G. Cui, S. Guo, L. Kong and X. Yang, "Human Target Detection Based on FCN for Through-Wall Radar Imaging," in *IEEE Geoscience and Remote Sensing Letters*, vol. 18, no. 9, pp. 1565-1569, Sept. 2021.

[3] H. Yan *et al.*, "Moving Targets Detection for Video SAR Surveillance Using Multilevel Attention Network Based on Shallow Feature Module," in *IEEE Transactions on Geoscience and Remote Sensing*, vol. 61, pp. 1-18, 2023.

[4] E. Chapin and C. W. Chen, "Airborne along-track interferometry for GMTI," in *IEEE Aerospace and Electronic Systems Magazine*, vol. 24, no. 5, pp. 13-18, May 2009.

[5] C. Mueh and M. Labitt, "Displaced-phase-center antenna technique," in *Lincoln Laboratory Journal*, vol. 12, no. 2, pp. 281-296, 2000.

[6] Ender and Joachim HG, "Space-time processing for multichannel synthetic aperture radar," in *Electronics & Communication Engineering Journal*, vol. 10, no. 1, pp. 29-38, 1999.

[7] L. Yang, T. Wang and Z. Bao, "Ground Moving Target Indication Using an InSAR System With a Hybrid Baseline," in *IEEE Geoscience and Remote Sensing Letters*, vol. 5, no. 3, pp. 373-377, July 2008.

[8] D. Cerutti-Maori and I. Sikaneta, "A Generalization of DPCA Processing for Multichannel SAR/GMTI Radars," in *IEEE Transactions on Geoscience and Remote Sensing*, vol. 51, no. 1, pp. 560-572, Jan. 2013.

[9] X. Wang, G. Gao, and S. Zhou, "Performance comparison and assessment of displaced phase center antenna and along-track interferometry techniques used in synthetic aperture radar-ground moving target indication," in *Journal of Applied Remote Sensing*, vol. 8, no.1, pp. 083504-1-083504-18, 2014.

[10] Z. Wang, Y. Wang, M. Xing, G. -C. Sun, S. Zhang and J. Xiang, "A Novel Two-Step Scheme Based on Joint GO-DPCA and Local STAP in Image Domain for Multichannel SAR-GMTI," in *IEEE Journal of Selected Topics in Applied Earth Observations and Remote Sensing*, vol. 14, pp. 8259-8272, 2021.

[11] D. Fu, J. Wen, J. Xu, G. Liao and S. Ouyang, "STAP-Based Airborne Radar System for Maneuvering Target Detection," in *IEEE Access*, vol. 7, pp. 62071-62079, 2019.

[12] X. Yang, Y. Sun, T. Zeng, T. Long and T. K. Sarkar, "Fast STAP Method Based on PAST with Sparse Constraint for Airborne Phased Array Radar," in *IEEE Transactions on Signal Processing*, vol. 64, no. 17, pp. 4550-4561, 1 Sept.1, 2016.

[13] J. Candès, X. Li, and Y. Ma, "Robust principal component analysis," in *Journal of the ACM*, vol. 58, no. 3, pp. 1-37, 2011.

[14] X. Zhang, D. Wu, D. Zhu and H. Zhou, "Multichannel SAR Moving Target Detection via RPCA-Net," in *IEEE Geoscience and Remote Sensing Letters*, vol. 19, pp. 1-5, 2022.

[15] Y. Guo, G. Liao, J. Li and T. Gu, "A Clutter Suppression Method Based on NSS-RPCA in Heterogeneous Environments for SAR-GMTI," in *IEEE Transactions on Geoscience and Remote Sensing*, vol. 58, no. 8, pp. 5880-5891, Aug. 2020.

[16] Y. Guo, G. Liao, J. Li and X. Chen, "A Novel Moving Target Detection Method Based on RPCA for SAR Systems," in *IEEE Transactions on Geoscience and Remote Sensing*, vol. 58, no. 9, pp. 6677-6690, Sept. 2020.

[17] F. H. C. Tivive and A. Bouzerdoum, "Toward Moving Target Detection in Through-the-Wall Radar Imaging," in *IEEE Transactions on Geoscience and Remote Sensing*, vol. 59, no. 3, pp. 2028-2040, March 2021.

[18] S. Zhang, X. Lu, J. Yu, Z. Dai, W. Su and H. Gu, "Clutter Suppression for Radar via Deep Joint Sparse Recovery Network," in *IEEE Geoscience and Remote Sensing Letters*, vol. 21, pp. 1-5, 2024.

[19] L. Wang *et al.*, "Ground Moving Target Indication and Relocation in Spaceborne MIMO-SAR Systems," in *IEEE Transactions on Geoscience and Remote Sensing*, vol. 61, pp. 1-25, 2023.

[20] J. Shan, Y. Zhang, Q. Zhao and J. Lin, "A K-means Clustering and Triangulation-Based Scheme for Accurate Detection of Multiple Adjacent Through-the-Wall Human Targets," in *IEEE Transactions on Instrumentation and Measurement*, vol. 72, pp. 1-13, 2023.

[21] S. Chen, M. Klemp, J. Taghia, U. Kühnau, N. Pohl and R. Martin, "Improved Target Detection Through DNN-Based Multi-Channel Interference Mitigation in Automotive Radar," in *IEEE Transactions on Radar Systems*, vol. 1, pp. 75-89, 2023.

[22] D. -M. Wu and J. -F. Kiang, "Dual-Channel Airborne SAR Imaging of Ground Moving Targets on Perturbed Platform," in *IEEE Transactions on Geoscience and Remote Sensing*, vol. 61, pp. 1-14, 2023.

[23] P. Huang *et al.*, "Imaging and Relocation for Extended Ground Moving Targets in Multichannel SAR-GMTI Systems," in *IEEE Transactions on Geoscience and Remote Sensing*, vol. 60, pp. 1-24, 2022.

[24] B. Ge, C. Fan, D. An, W. Wang, Z. Zhou and M. Chen, "A Novel Phase Calibration Method for Dual-Channel CSAR-GMTI Processing," in *IEEE Geoscience and Remote Sensing Letters*, vol. 17, no. 4, pp. 636-640, April 2020.

[25] C. Warren, A. Giannopoulos, and I. Giannakis, "gprmax: Open source software to simulate electromagnetic wave propagation for ground penetrating radar," Comput. Phys. Commun., vol. 209, pp. 163–170,2016.

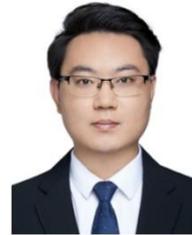

**Xiaodong Qu** Xiaodong Qu (Member, IEEE), associate researcher, received the B.S. degree from Xidian University, Xi'an, China, in 2012, and the Ph.D. degree from the University of Chinese Academy of Sciences, Beijing, China, in 2017.

His research interests mainly include array signal processing, through-the-wall radar imaging.

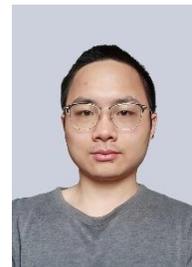

**Xiaolong Sun** Xiaolong Sun received the B.S. degree in Beijing Institute of Technology (BIT) in 2022. He is currently pursuing his M.S. degree at the Research Lab of Radar Technology, BIT.

His research interests mainly focus on through-the-wall moving targets detection by radar.

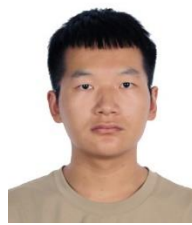

**Feiyang Liu** Feiyang Liu received the B.S. degree in electronic information engineering from Dalian University of Technology (DUT), Dalian, China, in 2023. He is currently pursuing the M.S. degree in the School of Information and Electronics at the Beijing Institute of Technology (BIT).

His research interests include unmanned aerial vehicle through-the-wall radar moving target detection and localization.

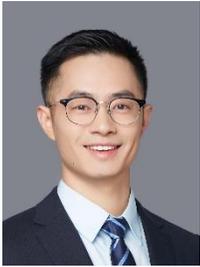

**Shichao Zhong** Shichao Zhong received the bachelor's degree in Geophysics from Chengdu University of Technology, Chengdu, China, in 2016, the Ph.D. degree at the Institute of Geology and Geophysics, Chinese Academy of Sciences, Beijing, China, in 2021.

He is currently a Post-Doctoral Associate with the School of Information and Electronics, Beijing Institute of Technology. His research interests include the parameter inversion, imaging, and hardware system development of ground-penetrating radar (GPR) and through-the-wall radar (TWR).

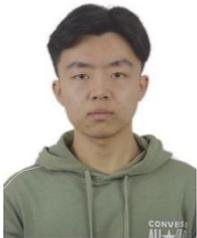

**Hao Zhang** Hao Zhang received the B.S. degree from the North China Electric Power University, Baoding, China, in 2022. He is currently working toward the M.S. degree at the School of Information and Electronics, Beijing Institute of Technology (BIT), Beijing, China.

His current research interests are in the areas of moving target localization with unmanned aerial vehicle based through-the-wall radar.

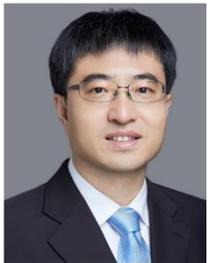

**Xiaopeng Yang** Xiaopeng Yang (Senior Member, IEEE), professor, received B.E. and M.E. degrees from Xidian University, China, in 1999 and 2002, respectively, and Ph.D. degree from Tohoku University, Japan, in 2007. He was a Post-Doctoral Research Fellow with Tohoku University from 2007 to 2008 and a Research Associate with Syracuse University, USA, from 2008 to 2010. Since 2010, he has been working with the School of Information and Electronics, Beijing Institute of Technology (BIT), Beijing, China, where he is currently a Full Professor.

His current research interests include the phase array radar, ground penetrating radar and through-the-wall radar.